\documentclass[aps,pra]{revtex4}
\usepackage{amssymb}
\usepackage{epsfig}
\usepackage{bm}

\def\be{\begin{equation}}
\def\ee{\end{equation}}
\def\bea{\begin{eqnarray}}
\def\eea{\end{eqnarray}}

\begin{document}

\title{New kind of condensation of Bose particles through stimulated
processes}
\author{Anatoly A. Svidzinsky$^{1,2}$, Luqi Yuan$^3$ and Marlan O. Scully$%
^{1,4}$}
\affiliation{$^1$Texas A\&M University, College Station, Texas 77843, USA; \\
$^2$Xiamen University of Technology, Xiamen 361024, China; \\
$^3$Shanghai Jiao Tong University, Shanghai 200240, China; \\
$^4$Baylor University, Waco, Texas 76798, USA }
\date{\today }

\begin{abstract}
We show that stimulated scattering of an isolated system of $N$ Bose
particles with initially broad energy distribution can yield condensation of
particles into excited collective state in which most of the bosons occupy
one or several modes. During condensation the total particle number and
energy are conserved, while entropy of the system grows. Onset of
condensation occurs at a critical particle occupation number when spectrum
narrowing due to stimulated processes overcomes spectrum broadening due to
diffusion. This differs from Bose-Einstein condensation in which particles
undergo condensation into the equilibrium state due to thermalization
processes.
\end{abstract}

\maketitle

\section{Introduction}

We are very pleased to dedicate this paper to the low-temperature leaders,
professors David Lee and John Reppy. They are low-temperature adventurers in
the spirit of C.T.\ Lane, the pioneer of the Yale superfluid He II group.
John's Bose-Einstein condensation (BEC) experiments in the porous vycor
glass \cite{Repp1,Repp2,Repp3} has been an inspiration to us in considering
fluctuations of BEC particle number in the canonical ensemble. When we had
the good fortune to convince David to come to work in Texas A\&M University
we were all very excited. Indeed it is an inspiration to see him (at 90
years young) still coming into the lab every day and publishing PRLs. As he
says: \textquotedblleft We've got a great story to tell. I just got to have
another PRL!\textquotedblright\ David's genius for asking deep questions has
enriched our lives in many ways. For example, his charming curiosity has led
us to a fruitful study of black hole entropy from a quantum optical
perspective \cite{Scul18}.

Bose-Einstein condensation implies macroscopic accumulation of particles on
the ground-state level (or in states other than the ground state, for
example, BEC with quantized vortices) of a Bose system at low temperature
and high density. It involves the formation of a collective quantum state
composed of identical particles. Superfluids and superconductors were, for a
long time, the only physical systems where the effect of BEC had been
observed. In the past decades, BEC has been theoretically predicted and
detected in other systems, including cold atomic gases \cite%
{Ande95,Davi95,Brad97}, collective modes and bosonic quasiparticles.
Examples of the latter are longitudinal electric modes \cite{Froh68b},
phonons \cite{Kaga07}, excitons \cite{Buto01}, polaritons \cite{Bali07},
exciton-polaritons \cite{Deng02,Kasp06}, photons \cite{Klae10}, rotons \cite%
{Meln11}, and magnons \cite{Boro84,Demo06,Math11,Vain15}. In these systems,
quasiparticles are externally pumped, but they are sufficiently long-lived,
so that their number $N$ is quasi-conserved. As a result, the chemical
potential $\mu =dE/dN$ is non-zero during the lifetime of the condensate.

BEC of photons has been observed in an optical microcavity \cite{Klae10}.
The cavity mirrors provide both a confining potential and a nonvanishing
effective photon mass, making the system formally equivalent to a
two-dimensional gas of trapped, massive bosons. Photon number conserving
thermalization was achieved by filling the cavity with a dye. The photons
thermalize to the temperature of the dye solution by multiple scattering
(absorption and re-emission) with the dye molecules. Upon increasing the
photon density, spontaneous onset of a macroscopic quantum phase with
massively populated ground-state mode on top of a broad thermal wing has
been detected \cite{Klae10}, which is a signature of the BEC transition.

An interesting phenomenon in open systems far from thermodynamic equilibrium
is the emergence of collective behaviors and self-organization, which is the
mechanism behind superfluorescence \cite{Boni75}, synchronization in
collective nonlinear dynamics \cite{Hein11,Zhan15,Shen20}, etc. In
biological systems, many theoretical works have suggested that the
collective behavior may have profound effects on the chemical and enzyme
kinetics \cite{Reim09}, and the cognitive function of brain \cite{Hame14}.
Among these works a widely used model is the Fr\"{o}hlich condensate \cite%
{Froh68,Froh68b,Froh70}. In 1968, Fr\"{o}hlich showed that the energy of a
driven set of oscillators would condense at the lowest vibrational mode once
the external energy supply exceeds a threshold \cite{Froh68,Froh68b}. The Fr%
\"{o}hlich condensate is a nonequilibrium phenomenon.

In this paper, we study a possibility of condensation of bosons into a
collective state with a macroscopic occupation of several modes with
different energy in a system with a nonlinear dynamics governed by
spontaneous and stimulated emission and absorption. Namely, we show that
such a system, apart from thermal-like states with a broad energy
distribution, can have steady-states in which one or several modes are
macroscopically occupied. In such condensed steady-states, there is exchange
of particles between the macroscopically occupied modes by means of
stimulated processes. The Bose system can condense into one of these
collective steady-states provided that for the initial particle distribution
the stimulated scattering dominates.

One should note that collective states into which particles condense are
different from fragmented BECs. Fragmented BEC occurs into a state which is
degenerate. Then two or more states are competing simultaneously for
Bose-Einstein condensation \cite{Muel06}. In our case, condensation occurs
into several modes with different energy.

\section{Model evolution equations for the particle distribution function}

\begin{figure}[t]
\vspace{0.5cm}
\par
\begin{center}
\epsfig{figure=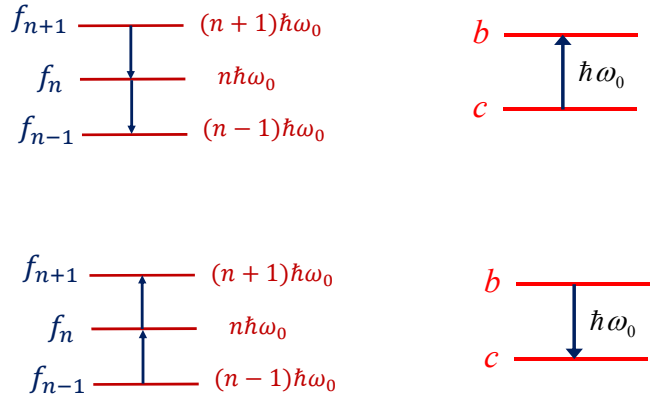, angle=270, width=10cm}
\end{center}
\caption{Probability flow diagram for photon frequency during excitation
(deexcitation) of the two-level system.}
\label{flow}
\end{figure}

We consider $N$ noninteracting bose particles trapped in a potential well
with equidistant energy levels $E_{n}=nE_{0}$. The system is isolated.
Distribution of particles over the levels is described by the distribution
function $f_{n}$, where $f_{n}$ is the average number of particles with
energy $E_{n}$. We turn on a $\delta -$function (in time) interaction with
an external system which makes particles jump into the adjacent lower energy
levels with some probability. We assume that during such event the
distribution function changes into 
\begin{equation}
f_{n}\rightarrow f_{n}+\kappa \left[ f_{n+1}(f_{n}+1)-f_{n}\left(
f_{n-1}+1\right) \right] ,  \label{n1}
\end{equation}%
where $\kappa $ is a dimensionless rate coefficient. Figure \ref{flow}
explains the probability flow of particles into the level $n$. Occupation
number of level $n$ increases because particles jump from the level $n+1$
into the level $n$ with a probability proportional to $f_{n+1}(f_{n}+1)$ and
decreases because particles go from the level $n$ into the level $n-1$ with
a probability $\propto f_{n}\left( f_{n-1}+1\right) $. Bosonic stimulation
is included in the factors $f_{n}+1$ and $f_{n-1}+1$ respectively. Those
factors make the system's evolution nonlinear.

Then we turn on another $\delta -$function interaction which promotes
particles into the adjacent higher energy levels and results in the change
of the distribution function into%
\begin{equation}
f_{n}\rightarrow f_{n}+\kappa \left[ f_{n-1}(f_{n}+1)-f_{n}\left(
f_{n+1}+1\right) \right] .  \label{n2}
\end{equation}

During consecutive scattering events (\ref{n1}) and (\ref{n2}) the total
number of particles $N=\sum_{n}f_{n}$ and their net energy $%
E=E_{0}\sum_{n}nf_{n}$ are conserved.

Our model, for example, can describe photons in a microcavity which are
analogous to massive particles in a harmonic trap. Radiation with a broad
band spectrum is trapped inside an ideal cavity. Inside the cavity there is
a two-level system with energy spacing $E_{0}$. The system can inelastically
scatter light changing photon frequency by $\pm E_{0}/\hbar $. This is,
e.g., the case in Raman scattering. Raman scattering is a process in which
an atom or a molecule absorbs a photon by undergoing a transition to a
virtual state and then decays to a real state by emitting a photon of lower
(Stokes Raman scattering) or higher (anti-Stokes Raman scattering) energy 
\cite{Scul18a}. During such scattering an atom or a molecule gives to or
takes away energy from the field by going to a higher or a lower energy
state, while the number of photons is conserved. We assume that there is no
energy dissipation, that is if the two-level system acquires energy $E_{0}$
from the radiation field then it will give this energy back in the next
scattering event.

The question we are interested in here is how the distribution function
evolves if the consecutive scattering events (\ref{n1}) and (\ref{n2}) occur
many times. One would expect that the distribution function will evolve into 
$f_{n}=$const, which is a stationary state of Eqs. (\ref{n1}) and (\ref{n2}%
). However, we will show that nonlinearity due to bosonic stimulation can
yield condensation of most of the particles into a single level if the
initial particle occupation numbers are large. This is similar to the line
narrowing in lasers. Photon condensation occurs into localized stationary
states which we will discuss next.

\section{Localized stationary states}

Here we obtain stationary states of Eqs. (\ref{n1}) and (\ref{n2}) for which
distribution function $f_{n}$ is narrow (localized). According to Eq. (\ref%
{n1}), after scattering event in which the two-level system goes to the
excited state the distribution function becomes 
\begin{equation}
\tilde{f}_{n}=f_{n}+\kappa \left[ f_{n+1}(f_{n}+1)-f_{n}\left(
f_{n-1}+1\right) \right] .  \label{n3}
\end{equation}%
In the next scattering event the two-level system goes back to the ground
state and the photon distribution function, according to Eq. (\ref{n2}),
changes to%
\begin{equation}
\bar{f}_{n}=\tilde{f}_{n}+\kappa \left[ \tilde{f}_{n-1}(\tilde{f}_{n}+1)-%
\tilde{f}_{n}\left( \tilde{f}_{n+1}+1\right) \right] .  \label{n4}
\end{equation}%
For stationary state 
\begin{equation}
\bar{f}_{n}=f_{n}.
\end{equation}%
This equation has the following stationary solution%
\begin{equation}
f_{n}=\frac{1}{\kappa }-1,\quad f_{n+1}=1-\kappa ,  \label{n5}
\end{equation}%
and other occupation numbers are equal to zero. Total photon number in this
state is $1/\kappa -\kappa $.

After scattering event in which the two-level system goes to the excited
state the distribution (\ref{n5}) changes to 
\begin{equation}
f_{n}=\frac{1}{\kappa }-1,\quad f_{n-1}=1-\kappa .  \label{n6}
\end{equation}%
In the next scattering event the distribution changes back to (\ref{n5}).
One should note that for $\kappa \ll 1$ only one mode in the state (\ref{n5}%
) is macroscopically occupied: $f_{n}\gg 1$.

Another stationary solution is 
\begin{equation}
f_{n-1}=\frac{1}{2}\left( \frac{1}{\kappa }-1\right) ,\quad f_{n}=\frac{1}{%
\kappa }-1,\quad f_{n+1}=\frac{1}{2}\left( 1-\kappa \right) ,  \label{n7}
\end{equation}%
and the other occupation numbers are equal to zero. The total number of
photons in this state is $3/2\kappa -\kappa /2-1$.

After scattering event in which the two-level system goes to the excited
state, distribution (\ref{n7}) changes to%
\begin{equation}
\tilde{f}_{n-2}=\frac{1}{2}\left( 1-\kappa \right) ,\quad \tilde{f}_{n-1}=%
\frac{1}{\kappa }-1,\quad \tilde{f}_{n}=\frac{1}{2}\left( \frac{1}{\kappa }%
-1\right) .  \label{n8}
\end{equation}%
In the next scattering event the distribution changes back to (\ref{n7}).
For $\kappa \ll 1$ the two modes in the state (\ref{n7}) are macroscopically
occupied: $f_{n},$ $f_{n-1}\gg 1$.

In next sections we demonstrate numerically that evolution Eqs. (\ref{n1})
and (\ref{n2}) can yield condensation of photons with initially broad
spectrum into localized stationary states, and show importance of the
stimulated scattering for condensation.

\section{Photon condensation via stimulated Raman scattering}

Nonlinear processes can result in condensation of photons into the localized
stationary states. As a consequence, an initially broad spectrum can evolve
into essentially delta function distribution conserving the net photon
energy and the total photon number. As a demonstration, we consider a
Gaussian initial photon energy distribution of the form%
\begin{equation}
f_{n}=f_{0}\exp \left( -\frac{\left( n-n_{0}\right) ^{2}}{2\sigma ^{2}}%
\right) .  \label{n9}
\end{equation}%
The total number of photons in the spectrum is 
\begin{equation}
N=f_{0}\sum_{n=0}^{\infty }\exp \left( -\frac{\left( n-n_{0}\right) ^{2}}{%
2\sigma ^{2}}\right) \approx \sqrt{2\pi }f_{0}\sigma .  \label{n10}
\end{equation}

\begin{figure}[t]
\begin{center}
\epsfig{figure=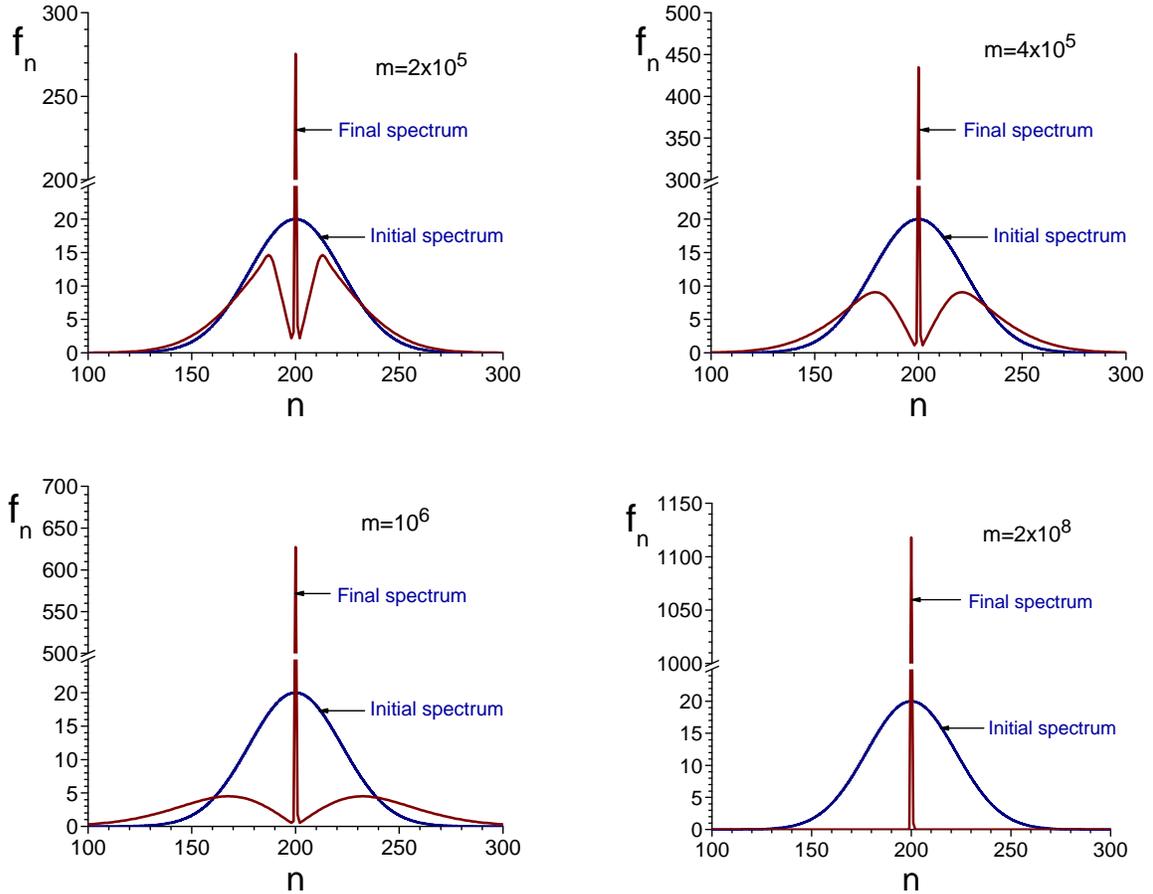, angle=270, width=18cm}
\end{center}
\caption{Initial and final radiation spectrum in the cavity. Initial
spectrum is given by the Gaussian distribution (\protect\ref{n9}) with $%
f_{0}=20$, $n_{0}=200$ and $\protect\sigma =22.36$. The final spectrum is
obtained by letting the system to evolve according to Eqs. (\protect\ref{n1}%
) and (\protect\ref{n2}) with $\protect\kappa =0.000892$ after different
number of scattering events $m=2\times 10^{5}$, $4\times 10^{5}$, $10^{6}$
and $2\times 10^{8}$.}
\label{condense}
\end{figure}

Next we take $f_{0}=20$, $n_{0}=200$ and $\sigma =\sqrt{500}=22.36$. Then
Eq. (\ref{n10}) yields that the total number of photons in the spectrum is $%
N=1121$. We want to condense these photons into a single localized state (%
\ref{n5}). Such localized state has $1/\kappa -\kappa $ photons. We take $%
\kappa =0.000892$, then $1/\kappa -\kappa =1121$, that is number of photons
in the localized state (\ref{n5}) is equal to the total number of photons in
the cavity. Under such conditions the initially broad spectrum with width $%
2\sigma =45$ (that is spectrum covers $45$ modes) condenses into a single
mode by means of the nonlinear Raman scattering mechanism.

To show this, we perform numerical simulations in which we let the initial
Gaussian spectrum evolve according to Eqs. (\ref{n1}) and (\ref{n2}) and
take the number of the scattering events to be $m=2\times 10^{5}$, $4\times
10^{5}$, $10^{6}$ and $2\times 10^{8}$. Figure \ref{condense} presents the
results. The initial broad spectrum gradually evolves into the single
localized state in which $f_{200}=1118.07$ and $f_{201}\approx 1$, while all
other $f_{n}$ are essentially equal to zero. The final localized state
contains $99.8$\% of the total field energy. The obtained numerically steady
state is well described by the analytical solution (\ref{n5}).

\begin{figure}[h]
\begin{center}
\epsfig{figure=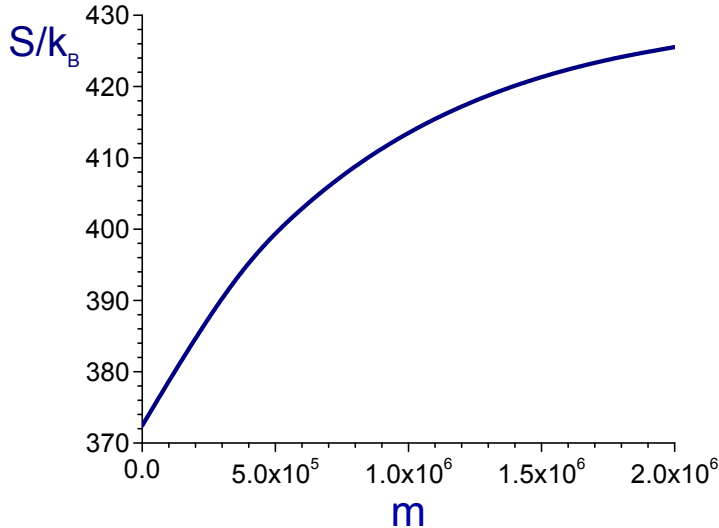, angle=270, width=10cm}
\end{center}
\caption{Entropy of the system of particles as a number of scattering events 
$m$ for spectrum evolution of Fig. \protect\ref{condense}.}
\label{entropy}
\end{figure}

According to the second law of thermodynamics, entropy $S$ of the isolated
particle system should grow during the condensation process. To show that
this is indeed the case we calculate $S$ as a function of the number of the
scattering events $m$. In terms of $f_{n}$ the entropy reads \cite{Land96}%
\begin{equation}
S=k_{B}\sum_{n}\left[ (1+f_{n})\ln (1+f_{n})-f_{n}\ln (f_{n})\right] .
\end{equation}

In Fig. \ref{entropy} we plot $S(m)$ for spectrum evolution of Fig. \ref%
{condense}. The total entropy of particles indeed increases with time even
though most of the particles condense into the same state. For $\kappa
=0.000892$, the entropy of the localized state (\ref{n5}) into which photons
are condensing is equal to $9.41k_{B}$, which is much smaller than the
entropy of the initial photon distribution $372k_{B}$ (see Fig. \ref{entropy}%
). Nevertheless, during condensation process the net entropy increases
because a small fraction of bosons undergoes very large spectral broadening
which overcomes the entropy decrease caused by the spectrum narrowing.

\begin{figure}[h]
\begin{center}
\epsfig{figure=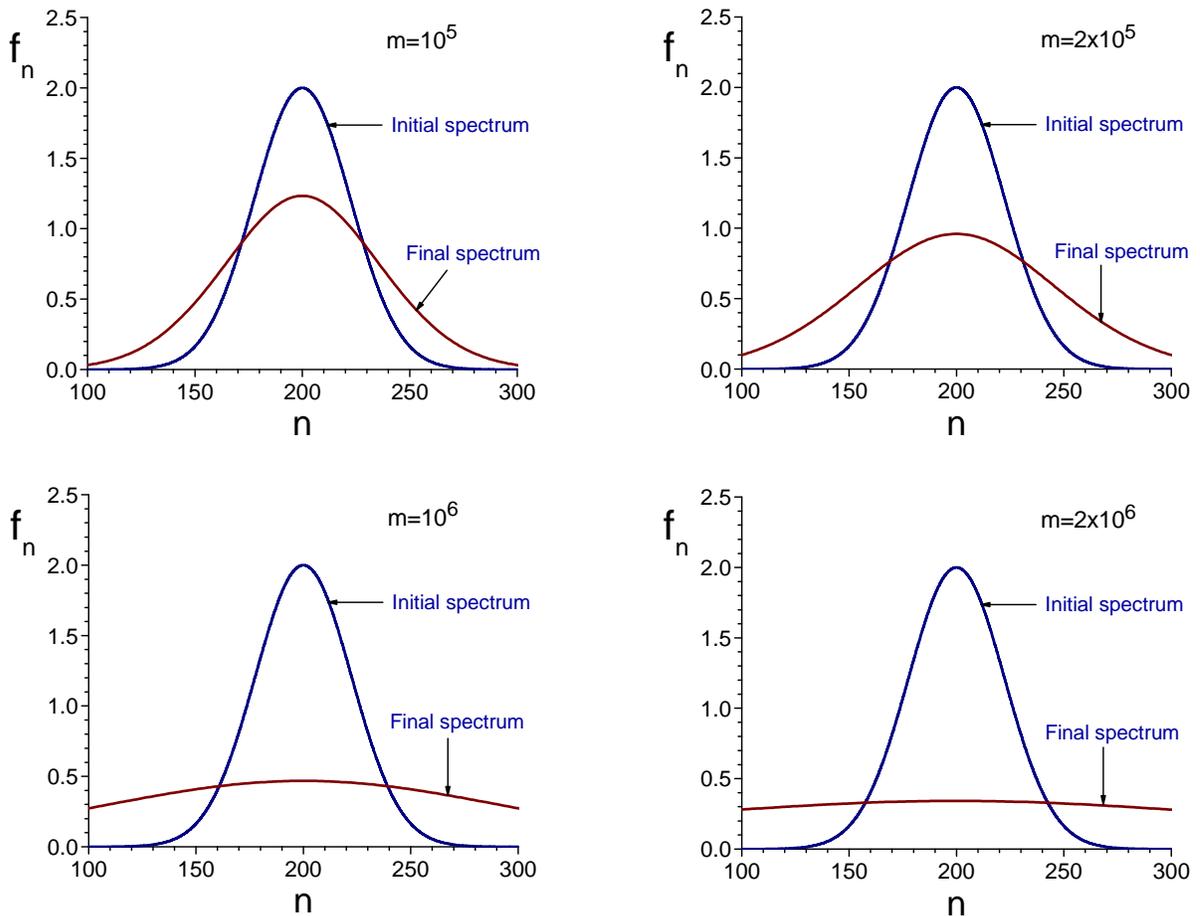, angle=270, width=18cm}
\end{center}
\caption{Initial and final radiation spectrum in the cavity. Initial
spectrum is given by the Gaussian distribution (\protect\ref{n9}) with $%
f_{0}=2$, $n_{0}=200$ and $\protect\sigma =22.36$. The final spectrum is
obtained by letting the system evolve according to Eqs. (\protect\ref{n1})
and (\protect\ref{n2}) with $\protect\kappa =0.00892$ after $10^{5}$, $%
2\times 10^{5}$, $10^{6}$ and $2\times 10^{6}$ scattering events. }
\label{spread}
\end{figure}

To demonstrate the crucial role of nonlinearity in spectrum narrowing
(stimulated scattering) we take the same initial Gaussian spectrum but now
with $f_{0}=2$. For such initial state the stimulated scattering is not
dominant. For $f_{0}=2$, $n_{0}=200$ and $\sigma =\sqrt{500}$ Eq. (\ref{n10}%
) gives that the total number of photons in the spectrum is $N=112$. To
match this number with the number of photons in the localized state (\ref{n5}%
) we take $\kappa =0.00892$, then $1/\kappa -\kappa =112$. Next we let the
initial Gaussian spectrum evolve according to Eqs. (\ref{n1}) and (\ref{n2})
and take the number of the scattering events $m=10^{5}$, $2\times 10^{5}$, $%
10^{6}$ and $2\times 10^{6}$. Figure \ref{spread} shows the result of
numerical simulation. Raman-like scattering in the linear regime yields
broadening of the initial spectrum. After a large number of scattering the
spectrum becomes constant which is another steady-state solution of Eqs. (%
\ref{n1}) and (\ref{n2}).

\section{Narrowing of spectrum with small photon occupation number using
high intensity seed}

\begin{figure}[h]
\begin{center}
\epsfig{figure=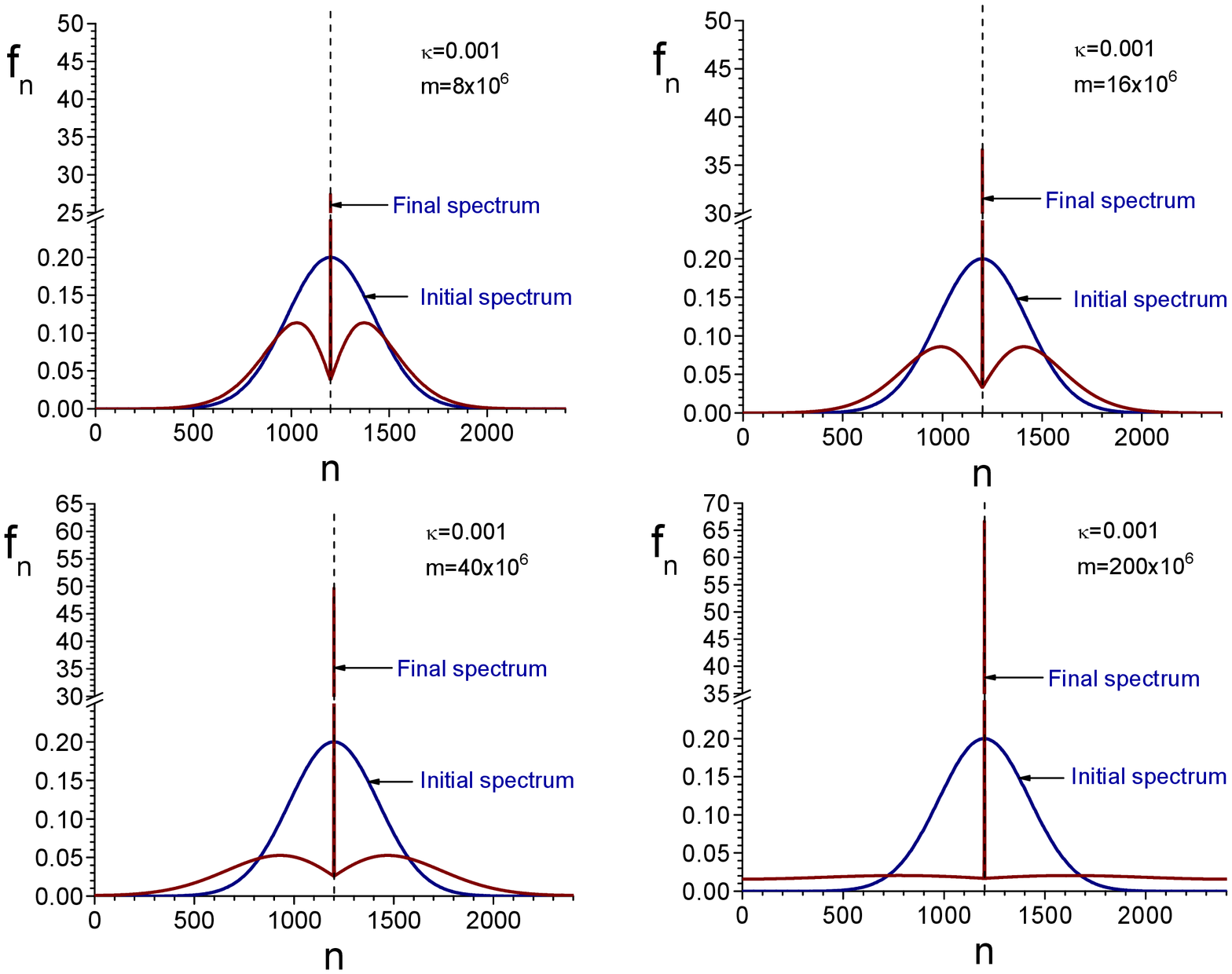, angle=270, width=18cm}
\end{center}
\caption{Initial and final radiation spectrum in the cavity. Initial
spectrum is given by the Gaussian distribution (\protect\ref{n9}) with $%
f_{0}=0.2$, $n_{0}=1200$ and $\protect\sigma =223.6$. In addition, there is
a $\protect\delta -$like seed pulse with $f_{1200}=900$ (shown by dash
line). The final spectrum is obtained by letting the system to evolve
according to Eqs. (\protect\ref{n1}) and (\protect\ref{n2}) with $\protect%
\kappa =0.001$ after different number of scattering events $m=8\times 10^{6}$%
, $16\times 10^{6}$, $40\times 10^{6}$ and $200\times 10^{6}$.}
\label{seed}
\end{figure}

Occupation number of photons near the maxima of Planck distribution is about 
$0.06$. For such low occupation numbers the Raman-like scattering yields
spectrum broadening. To condense the thermal solar spectrum into a narrow
frequency interval one can use a laser-like seed source which has a high
photon occupation number.

To demonstrate that this is possible, we take an initial spectrum given by
the Gaussian distribution (\ref{n9}) with $f_{0}=0.2$, $n_{0}=1200$ and $%
\sigma =\sqrt{50000}=223.6$. The total number of photons in the spectrum is $%
N=112$ and photon occupation number in each mode is $\leqslant 0.2$, which
alone yields no condensation. In addition, we add a $\delta -$like seed
pulse with $f_{1200}=900$ and let the system evolve according to Eqs. (\ref%
{n1}) and (\ref{n2}) with $\kappa =0.001$. Figure \ref{seed} shows the
spectrum after different number of the scattering events $m=8\times 10^{6}$, 
$16\times 10^{6}$, $40\times 10^{6}$ and $200\times 10^{6}$. After $%
200\times 10^{6}$ scattering events, $67$ (out of $112$) photons in the
initial broad spectrum (about $60\%$) condense into the seed mode, while the
rest of the spectrum broadens out. Thus, narrow-band sources with high
photon occupation number can be used for condensation of thermal radiation.

\section{Discussion and summary}

Steady-state of quantum systems can be non-thermal and have a narrow
spectral width. For example, if inverted atoms are injected sequentially
into a cavity, this can yield lasing in a cavity mode with the highest gain,
and the steady-state of light in the cavity is non-thermal \cite%
{Scul66,Scul67}. The laser linewidth can be very narrow, much narrower than
the natural linewidth of the atom's emission line \cite{Gord55}. In this
example, the energy flows from the atoms (active medium) into the cavity
field, which is balanced by the field leakage out of the cavity.

In this paper we consider a model in which photons in an ideal cavity
interact with a two-level atom. The atom can inelastically scatter light
changing the photon frequency by $\pm E_{0}/\hbar $ (Raman process). During
such scattering the number of photons is conserved. As the ground-state
two-level atom acquires energy $E_{0}$ from the radiation field, it becomes
excited and gives the energy $E_{0}$ back to the field by going to the
ground state in the next scattering event. Our system is analogous to a
system of $N$ Bose particles in the microcanonical ensemble.

We find that apart from a thermal-like steady-state with a broad energy
distribution, the system has steady-states in which one or several modes are
macroscopically occupied. The system can evolve into the state with the
narrow energy distribution if initially the stimulated scattering dominates.
Thus, Raman scattering, combined with photon stimulated emission and
absorption, can result in condensation of a broad-spectrum radiation into a
narrow frequency interval conserving the total photon number and energy. We
demonstrate the crucial role of stimulated processes (which make equations
nonlinear) in photon condensation and show that condensation can also occur
for weak (solar-like) radiation in the presence of intense laser beam acting
as a seed.

In our system, the spectrum narrowing occurs due to stimulated processes,
which is analogous to a laser. However, in contrast to the laser, there is
no need for an external energy source (active medium). The present effect
also differs from BEC because the latter occurs due to thermalization, with
the macroscopically populated mode being a consequence of equilibrium Bose
statistics.

Our findings could have practical applications in solar energy conversion
into electric power. Efficiency of semiconductor solar cells is low mostly
because they convert the broad solar spectrum into electric energy \cite%
{Wurf09,Henr80}. If it would be possible first to convert the broad spectrum
into a narrow one without changing the total radiation energy, this could
enhance the efficiency of a single pn junction cell by a factor of two \cite%
{Svid21}.

This work was supported by the Air Force Office of Scientific Research
(Grant No. FA9550-20-1-0366 DEF), the Office of Naval Research (Grants No.
N00014-20-1-2184), the Robert A. Welch Foundation (Grant No. A-1261), the
National Science Foundation (Grant No. PHY-2013771), the Natural Science
Foundation of Fujian (Grant No. 2021I0025), the Natural Science Foundation
of Shanghai (Grant No. 19ZR1475700), and the Fundamental Research Funds for
the Central Universities. L.Y. acknowledges support from the Program for
Professor of Special Appointment (Eastern Scholar) at Shanghai Institutions
of Higher Learning.

\end{document}